\documentclass{iaus}
\usepackage{graphicx}

\title[Star formation and maser surveys] 
{Investigating high-mass star formation through maser surveys}

\author[Ellingsen et al.]   
{S.P. Ellingsen$^1$, M.A. Voronkov$^2$, D.M. Cragg$^3$, A.M. Sobolev$^4$, \break
S.L. Breen$^1$, P.D. Godfrey$^3$}

\affiliation{$^1$School of Mathematics and Physics, University of Tasmania, Private Bag 21, Hobart 7001, TAS, Australia \break email: Simon.Ellingsen@utas.edu.au\\[\affilskip]
$^2$Australia Telescope National Facility CSIRO, PO Box 76, Epping 1710, NSW, Australia\\[\affilskip]
$^3$School of Chemistry, Building 23, Monash University, Victoria 3800, Australia\\[\affilskip]
$^4$Ural State University, Lenin ave. 51, 620083 Ekaterinburg, Russia}

\pubyear{2007}
\volume{242}  
\pagerange{??--??}
\date{?? and in revised form ??}
\jname{Astrophysical masers and their environments}
\editors{J. Chapman \& W. Baan, eds.}
\begin{document}

\maketitle

\begin{abstract}
  Interstellar masers are unique probes of the environments in which
  they arise.  In studies of high-mass star formation their primary
  function has been as signposts of these regions and they have been
  used as probes of the kinematics and physical conditions in only a
  few sources.  With a few notable exceptions, we know relatively
  little about the evolutionary phase the different maser species
  trace, nor their location with respect to other star formation
  tracers.  While detailed studies of a small number of maser regions
  can reveal much about them, other information can only be obtained
  through large, systematic searches.  In particular, such surveys are
  vital in efforts to determine an evolutionary sequence for the
  common maser species, and there is growing evidence that methanol
  masers may trace an earlier phase than the other common maser species
  of OH and water.  
\end{abstract}

\firstsection 
\section{Introduction}\label{sec:intro}

Interstellar masers are one of the best, if not {\em the} best
signpost of young high-mass star formation regions.  They are
relatively common, intense and because they arise at centimetre
wavelengths, are not affected by the high extinction that plagues
observations in other wavelength ranges.  The 6.7~GHz transition of
methanol has been found to be a particularly useful signpost as it is
both common and strong and appears to only be associated with
high-mass star formation regions (\cite{MENB03}).  The methanol
multibeam (MMB) survey currently being undertaken with the Parkes and
Jodrell Bank telescopes is using observations of the 6.7~GHz
transition of methanol to trace the Galactic distribution of star
formation and investigate the structure of our Galaxy (see the papers
by Green et al. and Pestalozzi et al. in these proceedings).

In addition to being excellent signposts of star formation regions,
masers have characteristics that make them very well suited to be
utilised as probes of the regions.  The maser emission is compact and
intense, which means that it can be studied at milliarcsecond
resolution using very long baseline interferometry (VLBI) techniques.
This makes masers ideal probes of the kinematics of the gas in which
they arise (e.g. \cite{T+05}).  They are potentially equally powerful
probes of the physical conditions in these same regions
(e.g. \cite{C+01}), however, the complexity of maser pumping schemes
provides significant challenges for robust interpretation of the
results.  Although progress has been slow towards the overall goal of
utilizing masers as tools to study star formation, the pace of advance
has recently accelerated.  The proliferation of complementary
high-resolution observations of star formation regions at millimetre
through mid-infrared wavelengths means that this trend is likely to
continue.

Some of the questions relating to masers are best addressed through
detailed studies of many transitions, molecules and wavelength bands
towards a small number of sources.  For example this is the best means
of determining where the different maser species lie with respect to
other objects within star formation regions.  However, favourite
sources which show strong emission from a wide range of maser
transitions are atypical of high-mass star formation regions and so
there are some questions for which a statistical approach is a better
way to proceed.  Uncommon phenomena must be associated with unusual
sources, or brief evolutionary phases.  What we really want to know
are the common properties of masers in star formation regions, as they
can give us an insight into the general characteristics of the
process.

\section{An Evolutionary Sequence for Masers}\label{sec:evolution}

There are four species of interstellar maser which are common in
high-mass star formation regions.  Mainline OH masers at 1.6~GHz and
water masers at 22~GHz have been studied since the 1960s, while the
1980s and 1990s saw the discovery of a number of strong transitions
from two different types of methanol masers.  The OH, water and
class~II methanol masers are found close to newly formed high-mass
stars, as indicated by the presence of H{\sc ii} regions and bright
infrared sources.  The class~I methanol masers are found towards
high-mass star formation regions, but are usually offset from the
high-mass protostars and other maser species.  The presence of an
interstellar maser within a star formation region signifies
``special'' physical conditions.  They are special in the sense that
masing is only predicted to occur for certain ranges of temperature,
density, molecular abundance etc.  Theoretical models find the most
common maser species - mainline OH, 6.7~GHz methanol and 22~GHz water
masers are inverted over a wide range of physical conditions
(e.g. \cite{CSG02}).

At present there remains significant uncertainty about where within
star formation regions the different maser species arise and the
evolutionary phase they are associated with.  However, as relatively
easily observed signposts of star formation regions, it would be very
desirable to be able to use the presence or absence of the different
maser species as an evolutionary clock.  Towards many star formation
regions more than one (sometimes all) of the common maser species are
observed, which implies that there is significant overlap between the
evolutionary phases traced.  The complexity of the goal of finding an
evolutionary sequence for masers in high-mass star formation regions
should not be underestimated, indeed \cite[De Buizer et
al. (2005)]{DRTP05} suggest that the different maser species trace a
variety of stellar phenomena and are hence associated with multiple
evolutionary phases.  It is only through statistical studies of large
samples of maser sources that a definitive answer will be obtained.

There are a few facts that are commonly agreed for the common maser
species - in the words of former US Secretary of Defense Donald
Rumsfeld ``Things we know we know'' :
\begin{enumerate}
\item Water masers usually trace outflows (e.g. \cite{FPT92}).
\item Water masers trace a generally earlier evolutionary phase than OH 
  masers (\cite{FC89}).
\item Where OH and methanol masers are observed towards the same
  regions the clusters arise from the same general region
  (\cite{C97}), although there is no coincidence between the
  individual maser spots of the different species (\cite{M+92}).
\end{enumerate}
A number of independent lines of evidence suggest that in many cases
the class II methanol masers are associated with a very early stage of
the high-mass star formation process.  In particular, only a small
percentage are associated with ultra-compact H{\sc ii} regions
(\cite{P+98,W+98}), but all are associated with millimetre
(\cite{P+02}) or submillimetre sources (\cite{W+03}) and many are
embedded within infrared dark clouds (\cite{E06}).  They are
associated with sources which have a spectral energy distribution
consistent with them being deeply embedded (\cite{M+05}).  This raises
the obvious question, are some of the class~II methanol masers
associated with an earlier phase of high-mass star formation than
water masers?

\subsection{Water and Class~II methanol}

There have been a small number of comparative studies of water and
class~II methanol masers.  \cite[Sridharan et al. (2002)]{S+02}
searched for water and methanol masers towards an infrared-based
sample of candidate high-mass star formation regions and found that
the majority of sources which had associated masers showed emission
from both species.  While a search for water masers towards a sample
of methanol masers by \cite[Szymczak, Pillai \& Menten (2005)]{SPM05}
achieved a detection rate of approximately 50\%.  These studies
suggest that water and methanol masers are commonly observed towards
the same star formation regions, however, targeted searches inevitably
suffer potential biases arising from the selection criteria.  A recent
complete survey for water maser emission towards the G333.6-0.2 giant
molecular cloud (\cite{B+07}) complements previous unbiased surveys of
6.7~GHz methanol (\cite{E+96}) and main-line OH masers (\cite{CHG80})
in the same region.  For the first time we have been able to compare
in detail the distribution of these different maser species over an
area of more than a square degree.  Figure~3 of Breen et al. shows a
striking dichotomy between the water and methanol masers in the
region.  All the water masers lie very close to the main axis of the
cloud, as traced by the bright mid-infrared emission, while the
methanol masers generally lie near the interface between the stronger
and more diffuse emission.

That the water masers are generally projected against bright mid
infrared emission along the main axis of the region suggests that they
are associated with relatively evolved high-mass star formation.  The
presence of methanol masers offset from the brightest mid-infrared
emission and their absence along the main axis suggests that they may
be tracing a second epoch of star formation in the region, triggered
by energetic winds and outflows from the mid infrared bright sources.
The water maser search of Breen et al. was only sensitive to water
masers with a peak flux density stronger than 5~Jy and searched a
smaller fraction of the giant molecular cloud than the methanol and OH
searches, so it is likely that there are some weak water maser sources
in the region that have not yet been detected.  It is not possible to
draw general conclusions on the basis of observations of a single
giant molecular cloud, however, the results strongly suggest that
6.7~GHz methanol masers trace an earlier evolutionary phase of
high-mass star formation than do luminous water masers.

\subsection{Class~I and class~II methanol}

\begin{figure}
 \includegraphics[height=0.6\textwidth]{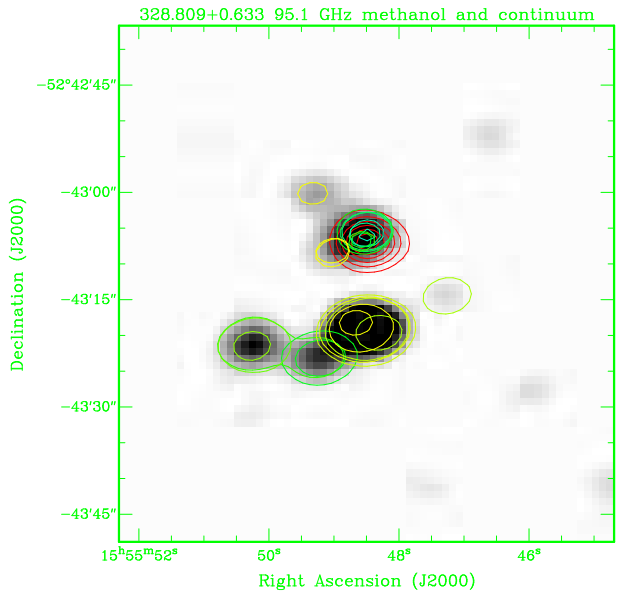}
 \center
 \caption{G328.81+0.63 shows strong (400~Jy) class~II methanol maser
   emission projected against the ultra-compact H{\sc ii} region (red
   contours - centred at approximately $\alpha$ = 15:55:48.5,
   $\delta$=-52:43:07).  The greyscale image shows the integrated
   95.1~GHz class~I methanol maser emission and the contours are a
   renzogram representing the velocity of the emission.}
 \label{fig:G328}
\end{figure}

Class~I methanol masers are the least well understood of the common
maser species.  They occur at the interface between outflows and the
ambient molecular cloud environment (\cite{PM90}), offset from the
other maser species.  One of the main difficulties in studying class~I
methanol masers has been the lack of good criteria on which to base
targeted searches.  \cite[Ellingsen (2005)]{E05} showed that
approximately 40\% of class~II methanol masers have an associated
class~I methanol maser within 30~arcseconds.  Recent ATCA observations
have shown that in many cases the class~I maser sources are not merely
chance detections towards other objects within the larger star
formation regions, but are intimately associated with the class~II
masers.  Figure~\ref{fig:G328} shows the 95.1~GHz class~I methanol
masers towards the well known class~II methanol maser source
G328.81+0.63.  The class~I masers are spread over 8 or more clusters
covering an extent of more than 30 arcseconds, the strongest emission
being to the south of the H{\sc ii} region.  In contrast the class~II
methanol masers lie within a single cluster projected against the
north-east of the H{\sc ii} region.  There is class~I methanol maser
emission from the same projected location as the class~II masers and
encompassing the same velocity range.  Theoretically strong class~I
and class~II methanol masers cannot coexist - the former are
collisionally excited, while the latter are radiatively pumped.  It
may be that the class~I maser are associated with an outflow directed
nearly perpendicular to the line of sight from the source pumping the
class~II masers.  Regardless of the detailed interpretation, these and
other ATCA observations demonstrate that in many cases class~I and
class~II methanol masers are associated with the same driving source,
rather than a coincidental association.

The presence of molecular outflows is thought to be one of the
earliest indications of the commencement of star formation and so we
might naively expect class~I masers to trace an even earlier
evolutionary phase than either water or class~II methanol
masers. \cite[Ellingsen (2006)]{E06} investigated the mid infrared
colours (from the GLIMPSE survey) of a sample of class~II methanol
the sample is relatively small, there is a strong tendency for the
sources with an associated class~I methanol maser to have redder
colours.  Redder colours are normally associated with more deeply
embedded, and hence younger sources.

\section{Conclusions}\label{sec:concl}

\begin{figure}
 \includegraphics[height=0.2\textwidth]{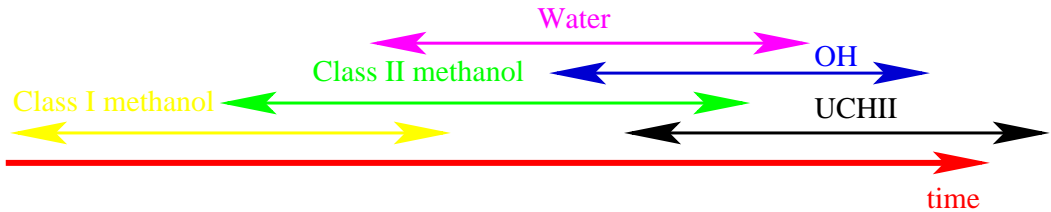}
 \caption{A ``straw man'' evolutionary sequence for masers in high-mass star
   formation regions.}
 \label{fig:sequence}
\end{figure}

Figure~\ref{fig:sequence} outlines a ``straw man'' evolutionary
sequence for masers in star formation regions that is consistent with
the results outlined above and previously established facts.  This
sequence appears consistent with the general properties of masers in
high-mass star formation regions, however, there are specific sources
which are not consistent with this sequence.  For example the
Turner-Welsh object near W3(OH) has associated water and OH maser
emission, but no methanol masers (\cite{A+03}), while IRAS16547-4247
has class~I methanol, water and OH, but no class~II methanol masers
(\cite{V+06}).  These may be rare sources, but only rigorous
statistical testing will show whether the proposed sequence usefully
describes the majority of cases.

\end{document}